\documentclass[pteplogo]{ptephy_v1}

\preprintnumber{XXXX-XXXX} 

\usepackage{amsmath} 
\usepackage{hyperref} 
\usepackage{graphics} 
\usepackage{subfig} 
\usepackage{url} 
\usepackage{multirow}

\usepackage{lineno}

\begin{document}

\title{Performance of new Kuraray wavelength-shifting fibers with short decay time}

\author{\name{\fname{Shoma}~\surname{Kodama}}{*}}
\author{\name{\fname{Hokuto}~\surname{Kobayashi}}{}}
\author{\name{\fname{Wataru}~\surname{Okinaga}}{}}
\author{\name{\fname{Kota}~\surname{Nakagiri}}{}}
\author{\name{\fname{Yasuhiro}~\surname{Nakajima}}{}}
\author{\name{\fname{Masashi}~\surname{Yokoyama}}{}}
\affil{Department of Physics, Graduate School of Science, The University of Tokyo, 7-3-1 Hongo, Bunkyo-ku, Tokyo, Japan \email{shoma@hep.phys.s.u-tokyo.ac.jp}}

\begin{abstract}%
We measure the decay time and the attenuation length of newly developed wavelength-shifting fibers, YS series from Kuraray, which have fast response.
Using a 405~nm laser, the decay times of the YS-2, 4, and 6 are measured to be 
$3.64 \pm 0.04$~ns,
$2.15 \pm 0.03$~ns, and
$1.47 \pm 0.02$~ns, respectively, 
for the light injection distance of 10~cm.
The decay time of Y-11 is measured to be 
$7.10 \pm 0.09$~ns
using the same system.
All fibers are found to have similar attenuation lengths of more than 4~meters.
When combined with the plastic scintillators EJ-200 and EJ-204, the YS series have better time resolution than Y-11, with light yields of 60--100\% of Y-11.
\end{abstract}

\subjectindex{H10, H15}

\maketitle

\section{Introduction}
Wavelength-shifting (WLS) fibers are widely used to collect light from scintillation detectors.
They are suitable to cover large areas but the time resolution is usually not very good.
If better time resolution is achieved, this technology can be applied to a much wider range of areas.
One of the non-negligible sources of time resolution in such systems is the emission decay time of the WLS fiber.
For example, the decay time of Kuraray's Y-11 WLS fiber, which is widely used in various experiments, is 7 ns (Table~\ref{tab:Fibers}), much longer than the typical rise time of plastic scintillators, $<1$ ns.

Recently, Kuraray has launched a new type of WLS fiber, the YS series~\cite{YSseries}.
A key feature of this new fiber is its short time constant for photon emission as shown in Table~\ref{tab:Fibers}.
A previous research reported promising results for YS-2~\cite{Alekseev:2021vbe}.
Since then, additional types of YS series have been released.

\begin{table}[tb]
   \centering
   \begin{tabular}{l|cccc}
       \hline
        & Y-11 & YS-2 & YS-4 & YS-6 \\
       \hline \hline
       Absorption peak (nm) & 430 & 422 & 420 & 414 \\
       Emission peak (nm)   & 476 & 474 & 470 & 462 \\
       Decay time (ns)      & 6.9 & 3.2 & 1.4 & 1.3 \\
       \hline
   \end{tabular}
   \caption{Characteristics of the Y-11 and YS series fibers provided by Kuraray~\cite{YSseries}.
   The decay time is measured using a small polystyrene plate with dye.}
   \label{tab:Fibers}
\end{table}

We report the measurement of performance of YS-2, 4, and 6 of the YS series.
For comparison, the Y-11 fiber is also measured.
All fibers are 1~mm in diameter and multi-clad type.
The YS series fibers are non-S-type, while the Y-11 fibers are S-type.
The core of the S-type has molecular orientation along the direction of the fiber.
The S-type fiber is stronger against clacking but the attenuation length is nearly 10\% shorter than the non-S-type.
The decay time and the attenuation length are measured by injecting light from a laser into the WLS fibers.
In addition, the light yield and the time resolution with plastic scintillators are measured with a 3~GeV/$c$ electron beam.


\section{Measurement of decay time and attenuation length}
\label{laser}
The decay time and the attenuation length of WLS fiber are measured by injecting light from a pulsed laser diode into the fiber.

\subsection{Measurement setup}

\begin{figure}[tb]
    \centering
    \includegraphics[width=12cm]{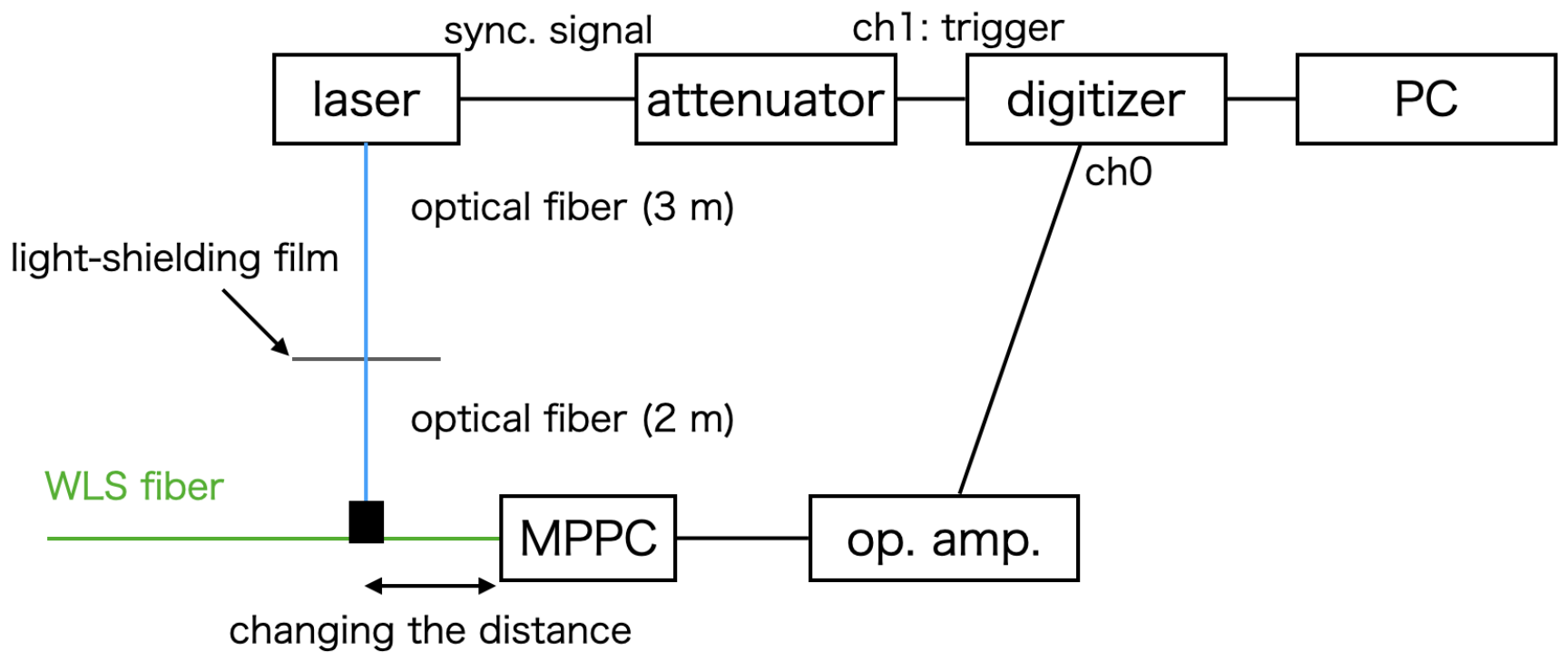}
    \caption{Schematic layout of the laser system.}
    \label{fig:SchematicDecaytime}
\end{figure}

A schematic view of the setup is shown in Fig.~\ref{fig:SchematicDecaytime}.
A 3~m long fiber is placed straight in a dark box.
One edge of the fiber is polished and connected to an MPPC (S13081-050CS, Hamamatsu Photonics).
The other edge is blackened with a marker pen to reduce the reflection.

We use the laser diode PLP-10 (Hamamatsu Photonics) as the light source.
The wavelength of the light is 405~nm, close to the absorption peak of the WLS fibers.
The width of the laser pulse is 60~ps.
The light intensity is reduced using light-shielding films.
The light is transported by optical fibers and is injected perpendicularly to the WLS fiber at a frequency of 10~kHz.
The distance between the light injection point and the MPPC is controlled with a jig to which an optical connector is attached.

The signal from the MPPC is amplified with an operational amplifier (Analog Devices AD8033) and recorded with the CAEN DT5730 14-bit, 500 MS/s 8-channel digitizer.
PLP-10 can send a TTL signal synchronized to the laser.
The TTL pulse is attenuated and also recorded and used as a trigger.

\subsection{Decay time}
For the measurement of the luminescent decay time of fibers, the light intensity is reduced to the single photon level.
The light is injected into the fiber at 10, 30, and 150 cm from the MPPC for cross-checks.

\begin{figure}[tb]
    \centering
    \subfloat[Waveform of a single-photon event]{
        \includegraphics[width = 6.5cm]{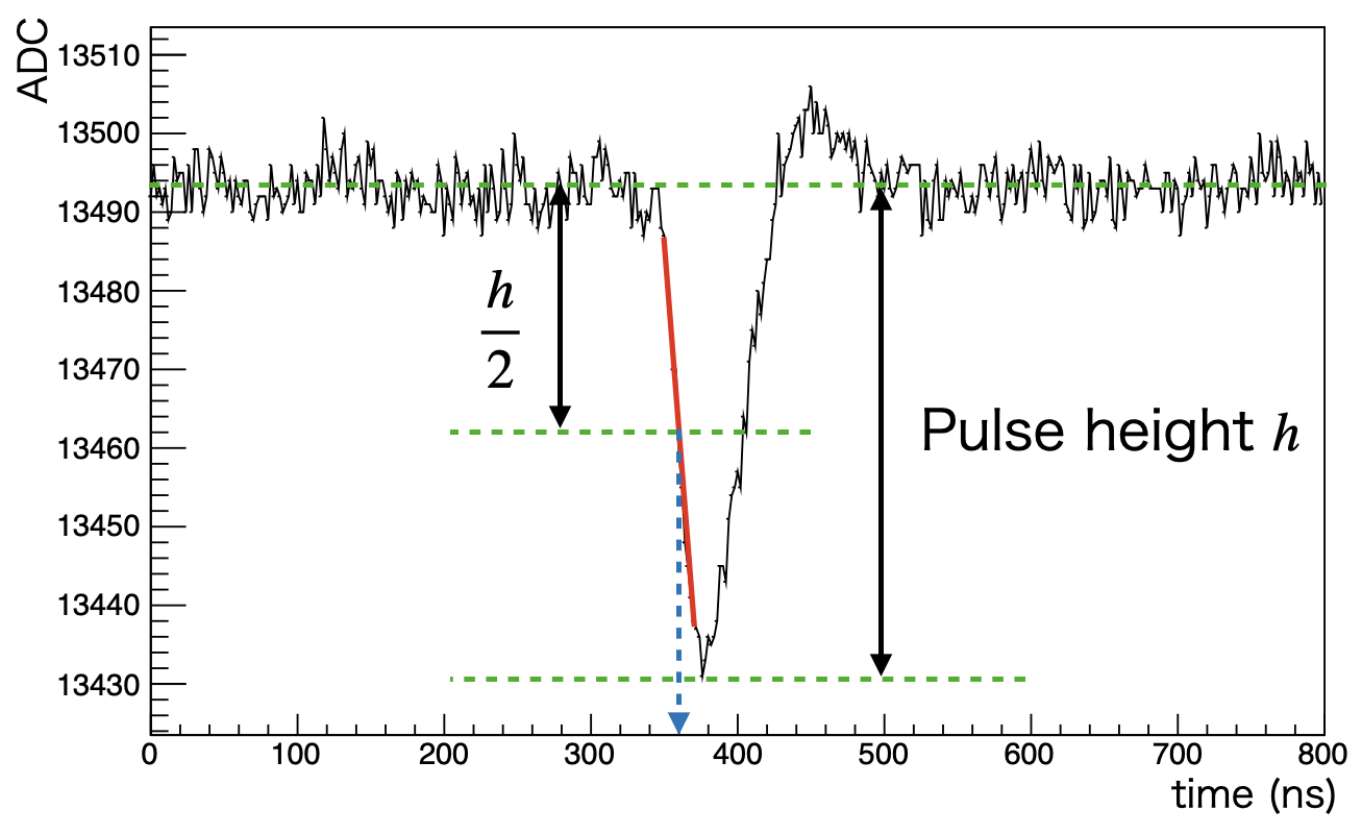}
        \label{fig:ADCDistribution_sub1}
    }
    \quad
    \subfloat[Pulse height and integrated ADC distribution for Y-11 at the distance of 10~cm]{
        \includegraphics[width = 6.5cm]{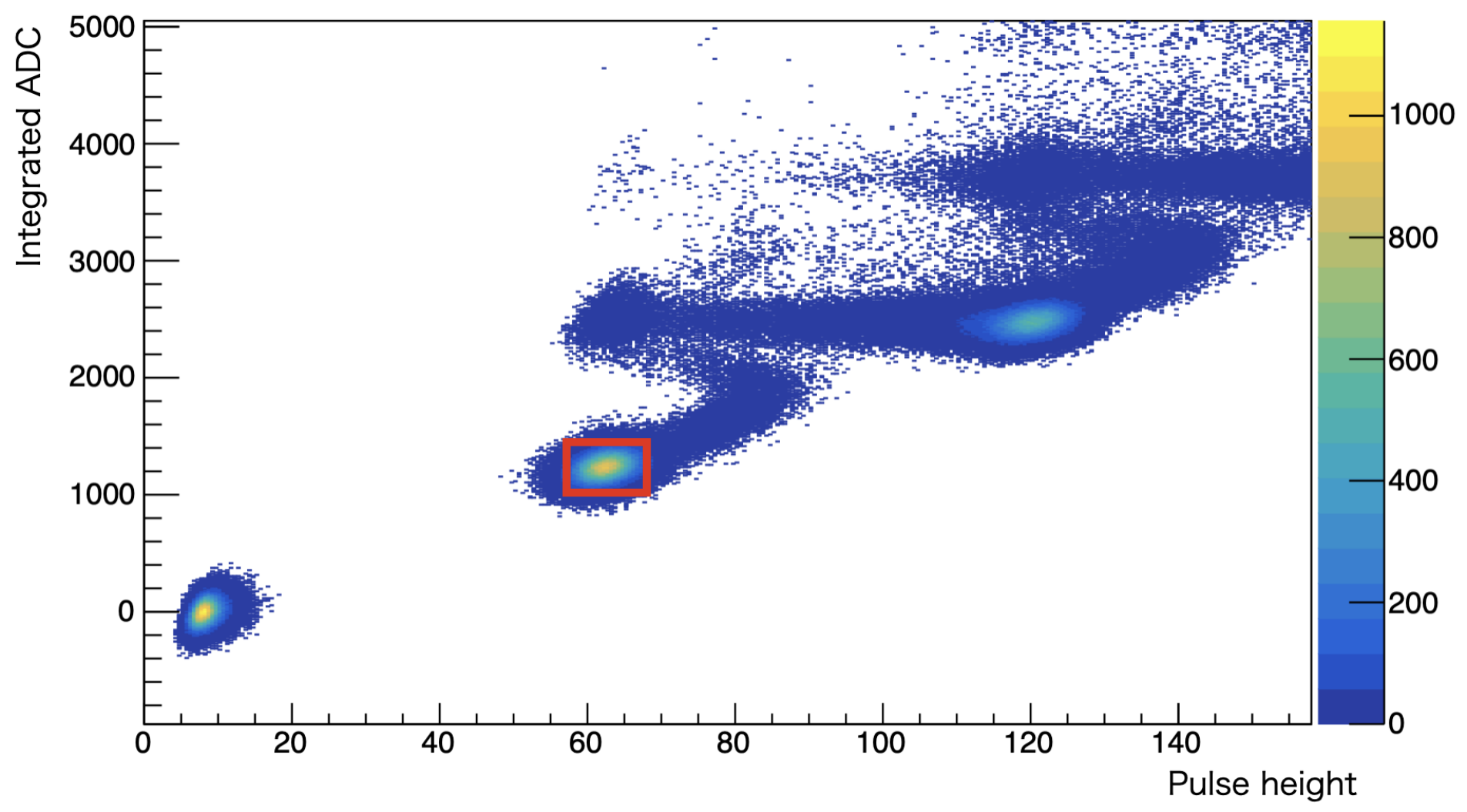}
        \label{fig:ADCDistribution_sub2}
    }
    \caption{Example of the waveform of the MPPC signal and two-dimensional signal distribution.}
    \label{fig:ADCDistribution}
\end{figure}

%
An example of the MPPC signal waveform is shown in Fig.~\ref{fig:ADCDistribution}\subref{fig:ADCDistribution_sub1}.
For each trigger, the waveform is recorded for 800~ns (400 points).
The pedestal of each event is calculated using the average of the first 100 points, while the remaining 300 points are used to measure the signal.
We use the pulse height and the integrated ADC value to determine the number of photons in an event.
The pulse height is the difference between the pedestal and the minimum ADC value.
The integrated ADC value is the sum of pedestal-subtracted ADC values of 300 points.
Figure~\ref{fig:ADCDistribution}\subref{fig:ADCDistribution_sub2} shows an example of the two-dimensional distribution of the pulse height and the integrated ADC value.
In addition to the peaks corresponding to the pedestal, single-photon, and double-photon events, there are contributions from effects such as after pulse and optical cross-talk of the MPPC.
We select events inside the red rectangular region in Fig.~\ref{fig:ADCDistribution}\subref{fig:ADCDistribution_sub2} as single-photon events.

We reject events that have hits in the first 200~ns used for the pedestal calculation by requiring the difference between the minimum and maximum ADC values in the region to be equal to or less than 30 ADC counts.
By this selection, 3.8\% of events are rejected.

The timing of the signal pulse is defined as the point at which the signal height is half the height of the peak in the falling edge.
Ten points around half the peak height of the pulse are fitted with a linear function to define the timing as shown in Fig.~\ref{fig:ADCDistribution}\subref{fig:ADCDistribution_sub1}.
The timing of the laser is determined using the synchronized signal to the laser.


The decay time is derived from the time difference between the laser pulse and the MPPC signal, $t$.
The signal distribution of $t$ is represented by a convolution of an exponential function with the decay time $\tau$ and a Gaussian with the time resolution for a single photon, $\sigma$.
The observed distribution of $t$ is fitted with the function
\begin{equation}
    N(t) = C \left[ 1 + \text{erf} \left( \frac{t - t_0 - \frac{\sigma^2}{\tau}}{\sqrt{2}\sigma} \right) \right] \cdot \exp \left( - \frac{t - t_0}{\tau} \right) + B,
    \label{eq:decayfitting}
\end{equation}
where $C$ is the normalization factor, $t_0$ is the time offset of the signal, and $B$ is the constant to account for the accidental background from the MPPC dark noise.
The upper limit of fitting range is determined so that the effect of the reflection from the other end of fiber is avoided.

\begin{figure}[tb]
    \centering
     \subfloat[10~cm]{
        \includegraphics[width = 6.5cm]{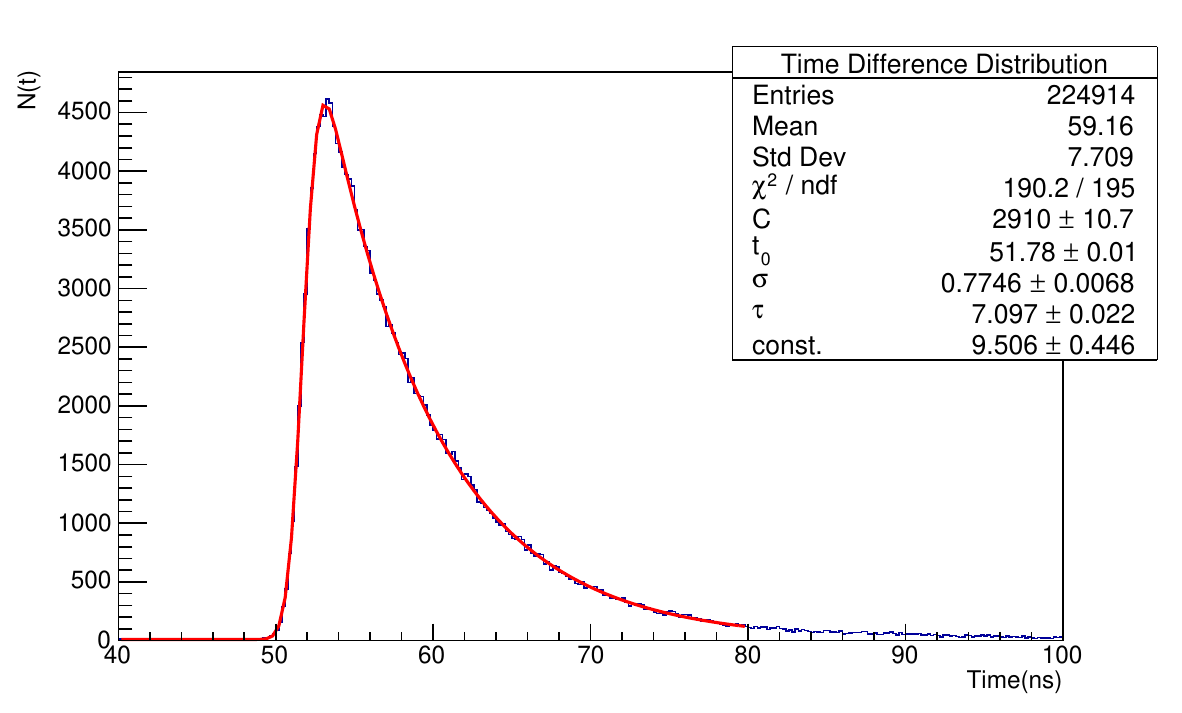}
        \label{fig:DecayTimeMeasurementResults_Y11}
    }
    \\
   \subfloat[30~cm]{
        \includegraphics[width = 6.5cm]{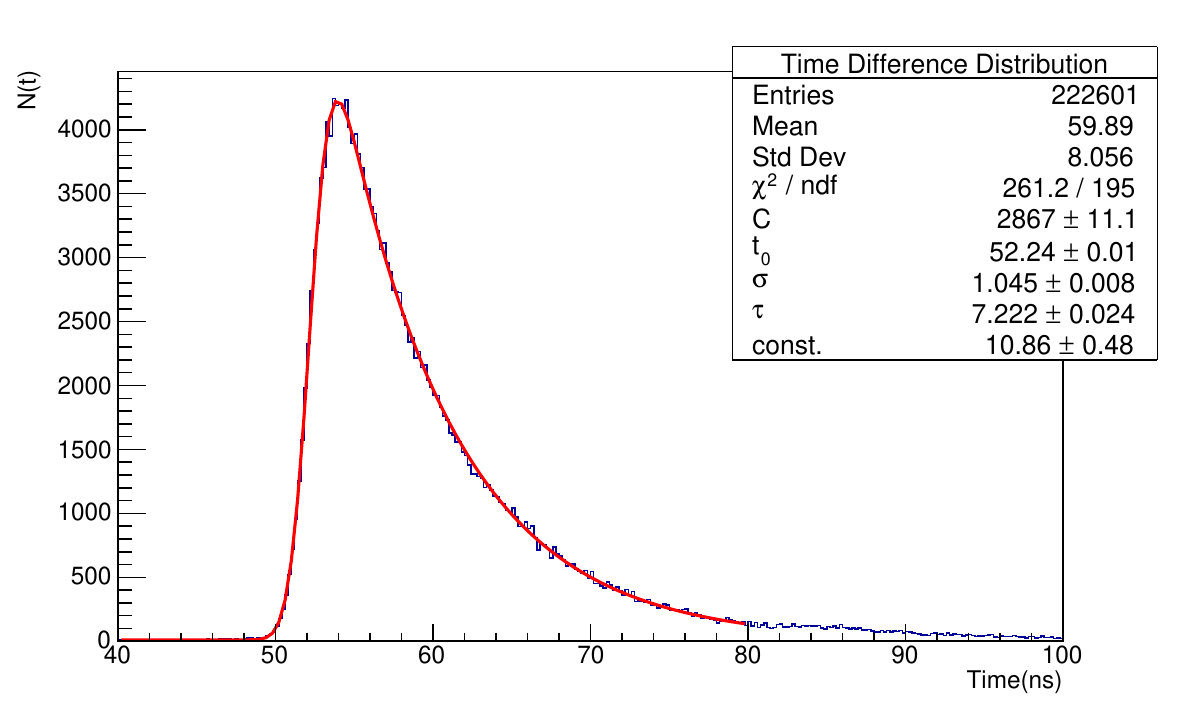}
        \label{fig:DecayTimeMeasurementResults_Y11_30cm}
    }
    \quad
    \subfloat[150~cm]{
        \includegraphics[width = 6.5cm]{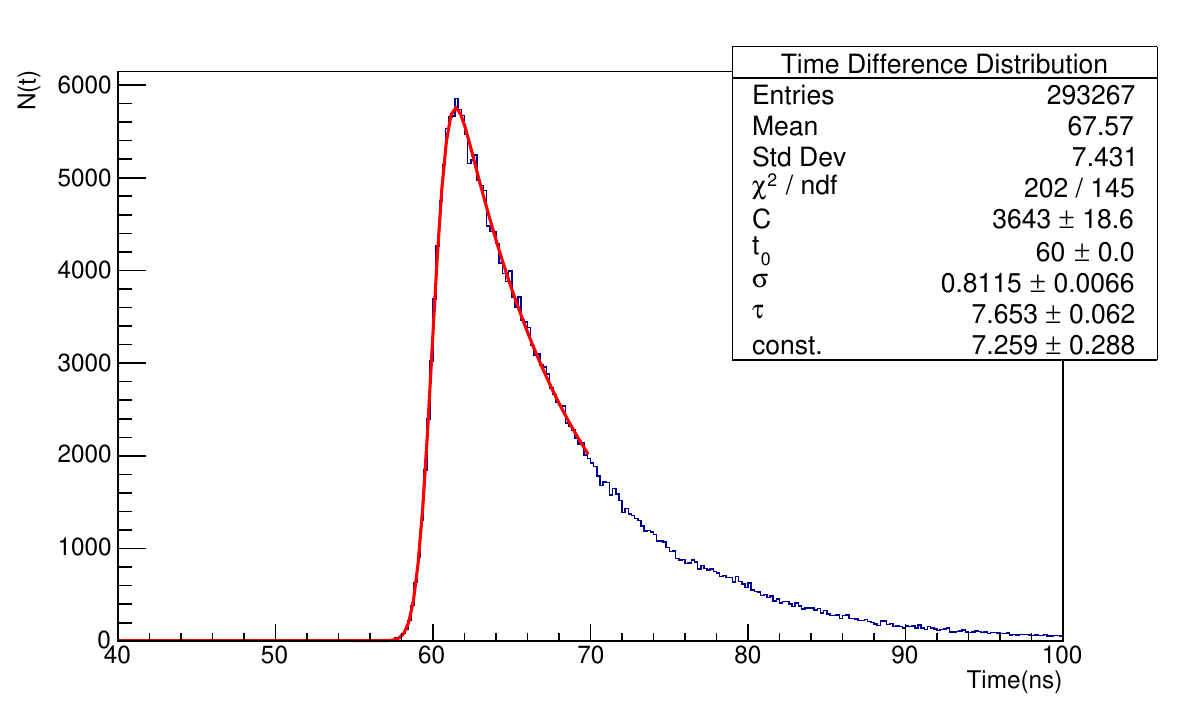}
        \label{fig:DecayTimeMeasurementResults_Y11_150cm}
    }
    \caption{Examples of time distributions and fit results with Y-11. The position of the light injection is 10~cm, 30~cm, and 150~cm from the MPPC. The fitting range is defined so that the effect of the reflection is avoided.}
    \label{fig:DecayTimeMeasurementPositionResults}
\end{figure}

We change the position of the light injection from 10 to 30 and 150~cm from the MPPC with Y-11 to check the effect of propagation length in the fiber.
Examples of the distributions are shown in Fig.~\ref{fig:DecayTimeMeasurementPositionResults}.
The values of decay time $\tau$ obtained from the fit are 7.10~ns, 7.22~ns, and 7.65~ns for the light injection distance of 10, 30, and 150~cm from the MPPC.
For a longer distance, the variation of propagation time in the fiber becomes larger and makes the apparent decay time longer.
We simulated the propagation of light in the fiber and confirmed that such effect exists.
In this paper, we use 10~cm as the nominal distance to report the decay time.

\begin{figure}[tb]
    \centering
    \subfloat[YS-2]{
        \includegraphics[width = 6.5cm]{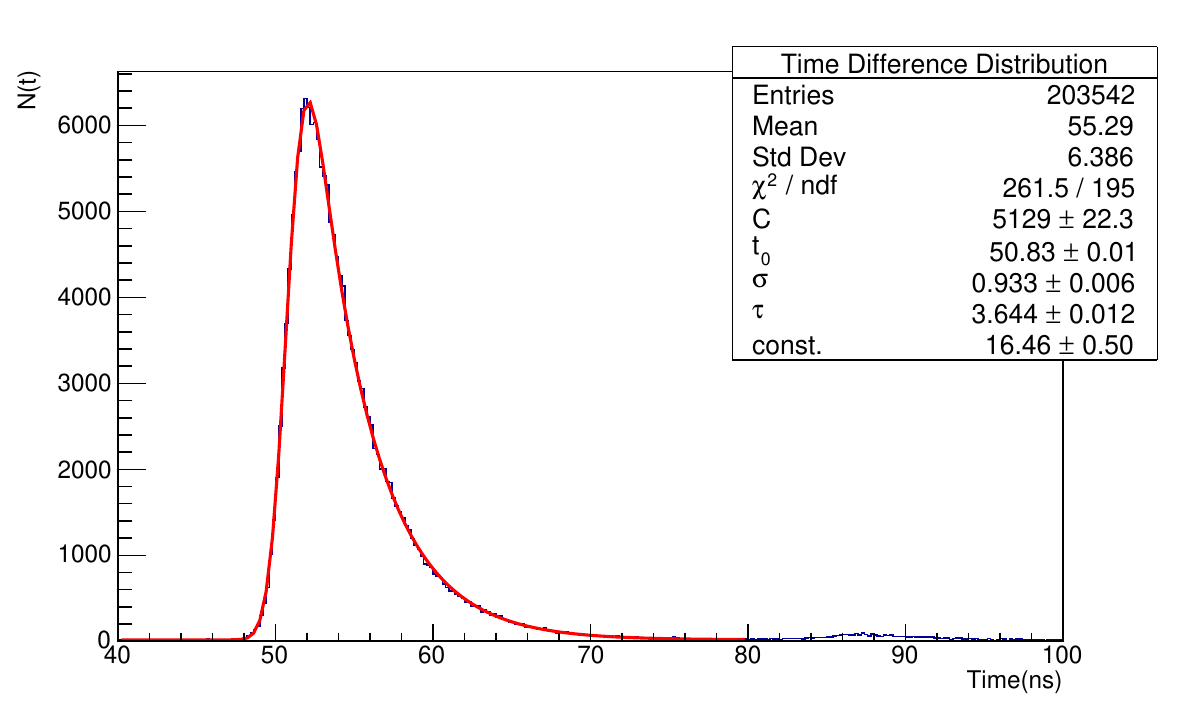}
        \label{fig:DecayTimeMeasurementResults_YS2}
    }
    \\
    \subfloat[YS-4]{
        \includegraphics[width = 6.5cm]{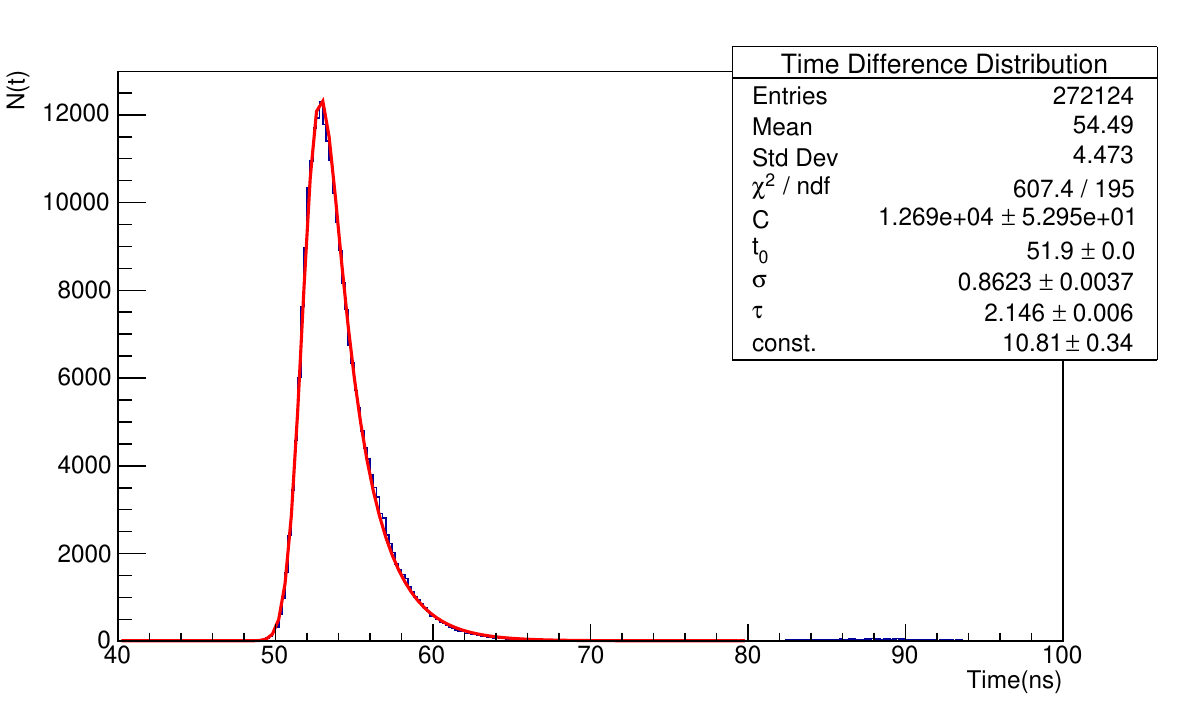}
        \label{fig:DecayTimeMeasurementResults_YS4}
    }
    \quad
    \subfloat[YS-6]{
        \includegraphics[width = 6.5cm]{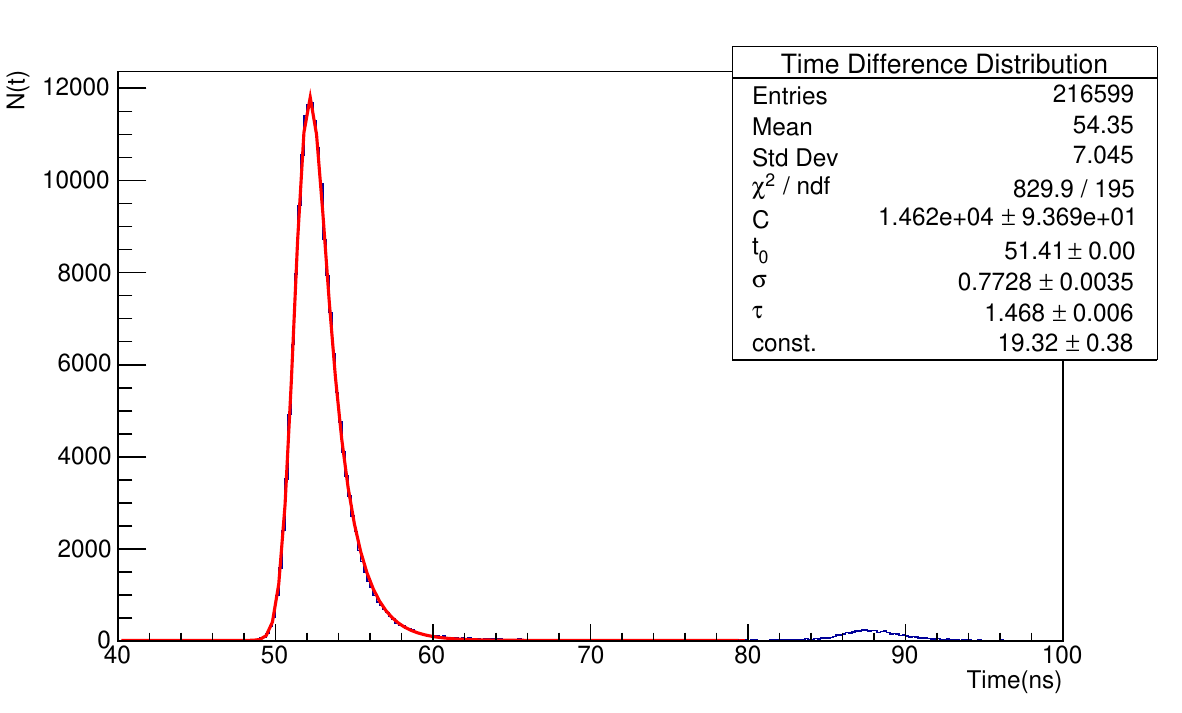}
        \label{fig:DecayTimeMeasurementResults_YS6}
    }
    \caption{The time distributions and fit results for YS-2, 4, and 6. The position of the light injection is 10~cm from the MPPC.
    The reflection is seen around 87~ns, which is consistent with the calculation from the lengths of cables and fiber.
    The fitting range is below 80~ns to avoid the effect of the reflection.}
    \label{fig:DecayTimeMeasurementResults}
\end{figure}

The measured $t$ distributions for YS-2, 4, and 6 at the distance of 10~cm are shown in Fig.~\ref{fig:DecayTimeMeasurementResults}.
The upper limit of fitting range is set to $t$=80~ns to avoid the effect of reflection.

Three sources of systematic uncertainties are considered.
The effect of the fitting range including the effect of the reflection is estimated by changing the upper limit of the fitting range from 70 to 100~ns.
The effect of time resolution of the measurement system is evaluated by fixing the resolution $\sigma$ in the fitting.
The value of $\sigma$ is evaluated by directly illuminating the MPPC with the laser to be 0.77~ns.
We assign half of the difference between with and without fixing $\sigma$ in the fitting as the uncertainty related to the time resolution of the measurement system.
To evaluate the effect of the injection position, the coupling of light injection, and the individual differences among fiber samples, we repeat the measurement with five different Y-11 fibers.
By taking half of the maximum and minimum, the uncertainty due to the measurement reproducibility is estimated to be 0.5\%.
The total uncertainties are estimated by adding all uncertainties in quadrature.

The measured decay times and their uncertainties are summarized in Table~\ref{tab:DecayTime_Results}.
We measure five fibers for Y-11 and the average of the measured values is shown.
For the YS series, we measure one fiber for each type.

\begin{table}
    \centering
    \begin{tabular}{c||c||cccc|c} \hline
             & & \multicolumn{5}{c}{Uncertainty} \\ \cline{3-7}
    Fiber    & Decay time & \multirow{2}{*}{Fitting} & Fit  & Measurement  & Repro-  &  Total \\ 
             & & & range & resolution & ducibility & uncertainty\\ \hline \hline
     Y-11    & 7.10 & 0.02 & 0.04 & 0.07 & 0.04 & 0.09 \\
     YS-2    & 3.64 & 0.01 & 0.02 & 0.03 & 0.02 & 0.04\\
     YS-4    & 2.15 & 0.01 & 0.01 & 0.02 & 0.01 & 0.03\\
     YS-6    & 1.47 & 0.01 & 0.01 & 0.02 & 0.01 & 0.02\\ 
     \hline
    \end{tabular}
    \caption{Measured decay time and estimated uncertainty of each fiber(ns).}
    \label{tab:DecayTime_Results}
\end{table}

The decay times of Y-11 and YS-2 are in agreement with the measurement in Ref.~\cite{Alekseev:2021vbe}.
The YS-4 and YS-6 have shorter decay times than YS-2.
The YS series is confirmed to have a faster response than Y-11.
These results indicate that the YS series is a good choice for the detector requiring a good time resolution.

\subsection{Attenuation length}


We model the light attenuation in the WLS fiber assuming two components with short and long attenuation lengths:
\begin{equation} \label{eq:attenuation}
    L = L_0 \cdot \left[
        \alpha \cdot \exp \left( - \frac{x}{A_L} \right) 
        + (1-\alpha) \cdot \exp \left( - \frac{x}{A_S} \right) 
    \right],
\end{equation}
where $L$ and $L_0$ are the attenuated and unattenuated light yields, $x$ is the distance between the light injection position and the MPPC, and $A_L$ and $A_S$ are the long and short attenuation lengths, respectively, and $\alpha$ is the fraction of the long attenuation component.

\begin{figure}[tb]
    \centering
    \includegraphics[width=10cm]{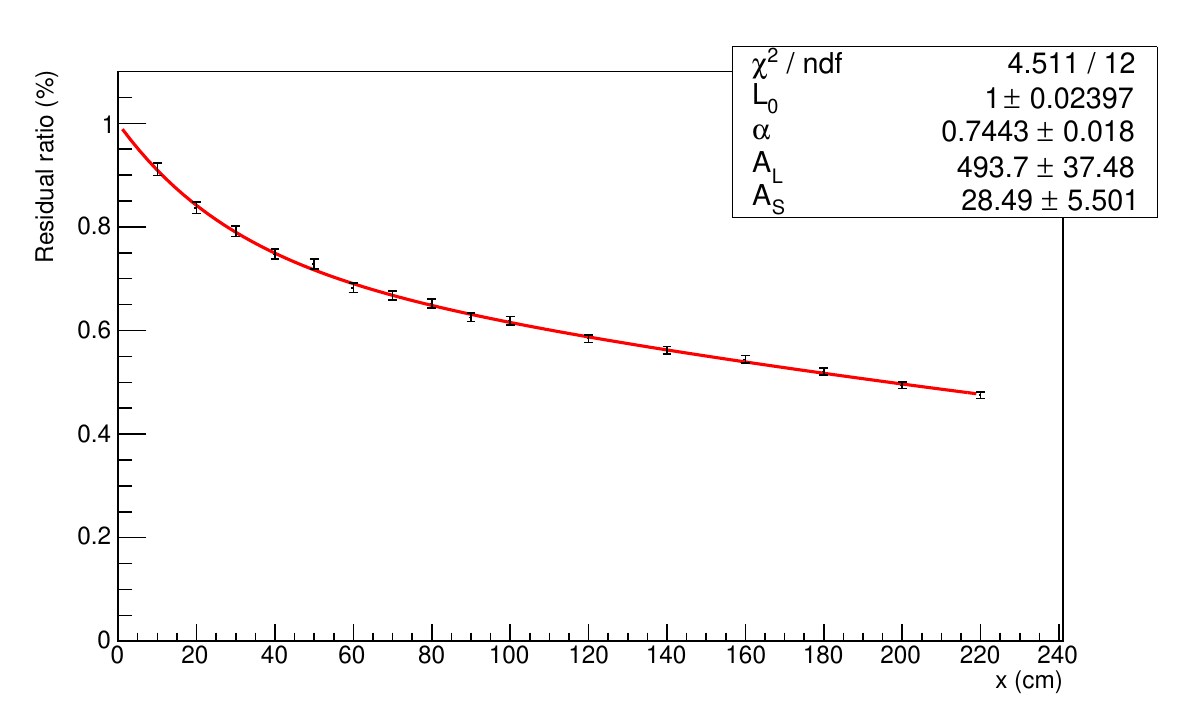}
    \caption{Example of the residual ratio after propagation of the photon in the fiber as a function of the distance between the MPPC and the light injection position.}
    \label{fig:AttenuationMeasurementResults}
\end{figure}

We measure the light yield as a function of $x$.
The range of $x$ is from 10~cm to 220~cm.
The uncertainty of the light yield at each point is estimated to be 1.4\%, dominated by the stability of the laser intensity and the optical coupling at the injection point to the fiber.
It is estimated by putting the same fiber in the same position every five minutes from 30 minutes to 180 minutes after turning on the laser and taking the standard deviation of the light yields normalized by the average light yield.
The measured light yield is fitted with the function in Eq.~\ref{eq:attenuation}.
An example of the fitting is shown in Fig.~\ref{fig:AttenuationMeasurementResults}.
It is scaled so that $L_0$ equals one.
The results are summarized in Table~\ref{tab:Attenuation_Results1}.
The attenuation lengths of YS series fibers are similar to that of Y-11 and longer than 4~m.
\begin{table}[tb]
    \centering
    \begin{tabular}{c|ccc}
        \hline
        Type & $\alpha$ & $A_L$ (cm) & $A_S$ (cm) \\
        \hline \hline
        Y-11 & $0.74 \pm 0.02$ & $494 \pm 37$ & $28 \pm 6$ \\
        YS-2 & $0.75 \pm 0.02$ & $437 \pm 19$ & $34 \pm 5$ \\
        YS-4 & $0.84 \pm 0.02$ & $558 \pm 27$ & $26 \pm 8$ \\
        YS-6 & $0.73 \pm 0.02$ & $527 \pm 22$ & $29 \pm 4$ \\
        \hline
    \end{tabular}
    \caption{Attenuation lengths measured using the laser. The uncertainties are from fitting.}
    \label{tab:Attenuation_Results1}
\end{table}

\section{Measurement of light yield and time resolution with plastic scintillators}
\label{beam}
To evaluate the performance of the WLS fibers combined with plastic scintillators, we measure the light yield and time resolution using a 3~GeV/$c$ electron beam at the KEK AR test beamline~\cite{Mitsuda:2023gba}.

\begin{figure}[tb]
    \centering
    \includegraphics[width=10cm]{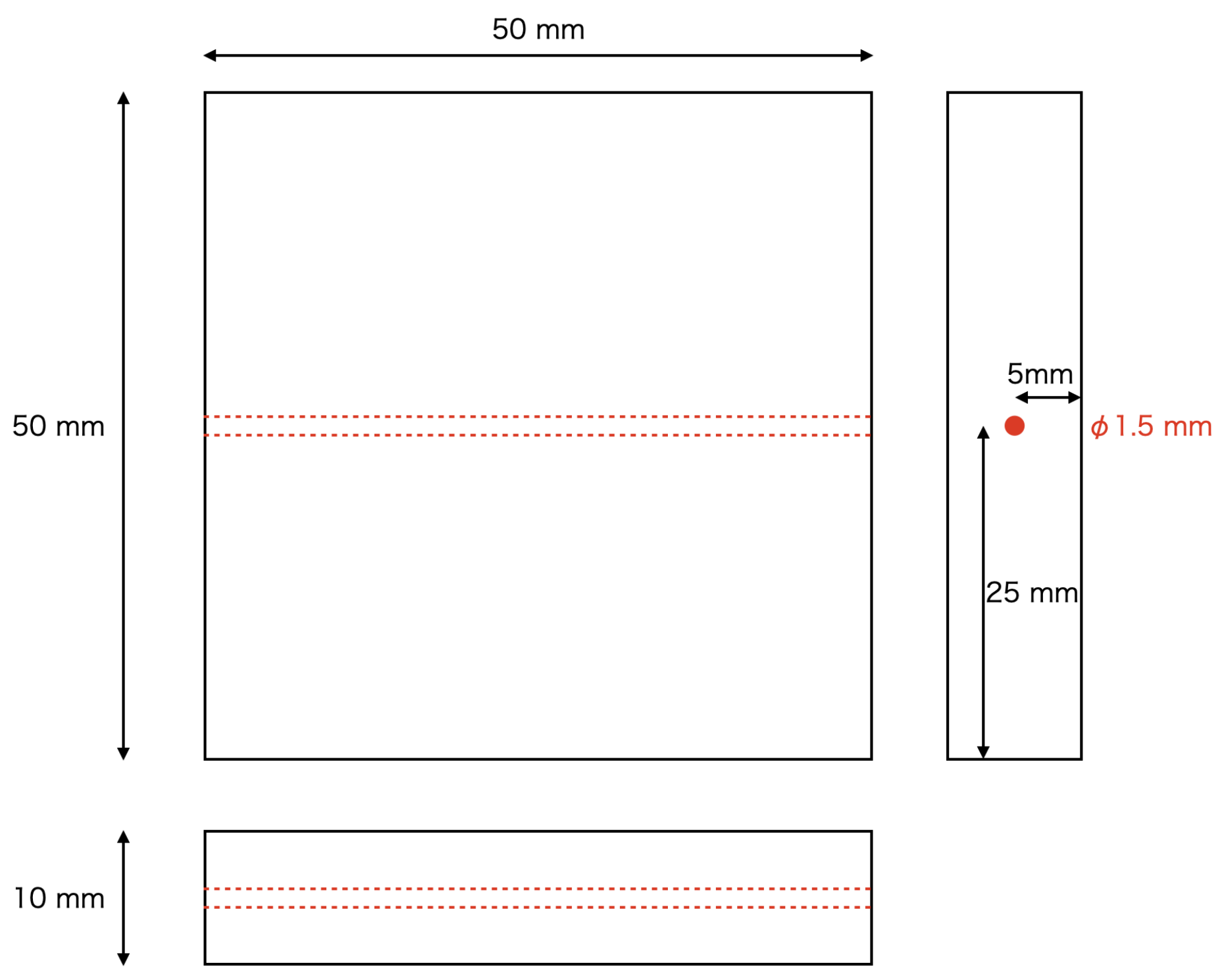}
    \caption{Dimension of the plastic scintillator.}
    \label{fig:ScintillatorDesign}
\end{figure}

We use EJ-200 and EJ-204 plastic scintillators by ELJEN~\cite{ELJEN}.
Basic characteristics are shown in Table~\ref{tab:Scintillators}.
The dimension of the scintillator is 5~cm $\times$ 5~cm $\times$ 1~cm.
There is a 1.5~mm diameter hole to insert a fiber as shown in Fig.~\ref{fig:ScintillatorDesign}.
We do not use any glue or grease and the fiber is optically coupled to the scintillator with only air. 
All the scintillators are wrapped with an aluminized polyester sheet and covered with a black sheet.

\begin{table}[tb]
   \centering
   \begin{tabular}{l|cc}
       \hline
        & EJ-200 & EJ-204 \\
       \hline \hline
       Emission peak (nm)   & 425 & 408 \\
       Rise time (ns)       & 0.9 & 0.7 \\
       Decay time (ns)      & 2.1 & 1.8 \\
       \hline
   \end{tabular}
   \caption{Characteristics of the EJ-200 and EJ-204 plastic scintillators provided by ELJEN~\cite{ELJEN}.}
   \label{tab:Scintillators}
\end{table}

\subsection{Measurement setup}

\begin{figure}[tb]
    \centering
    \includegraphics[width=10cm]{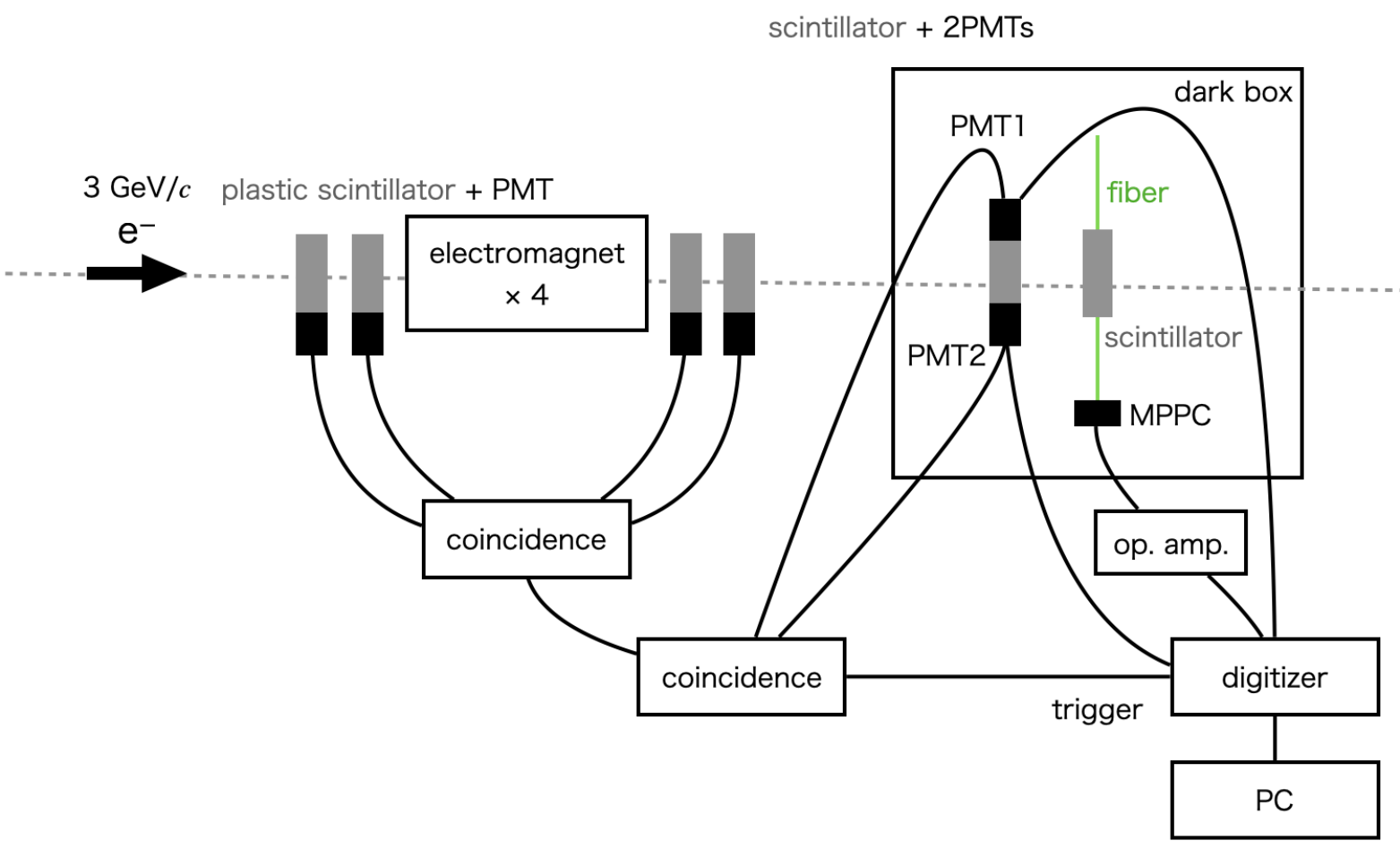}
    \caption{Schematic layout of measurement with an electron beam.}
    \label{fig:SchematicBeamtest1}
\end{figure}

A schematic view of the setup is shown in Fig.~\ref{fig:SchematicBeamtest1}.
The plastic scintillator, WLS fiber, and MPPC are placed in a dark box.
The distance between the center of the scintillator and MPPC is 30~cm.
The beam center is aligned to the center of the plastic scintillator.
The spacial profile of the electron beam is about 9~cm $\times$ 2~cm.
A trigger is issued by a coincidence of four scintillators placed in the beamline and two PMT signals from a trigger scintillator (4~cm $\times$ 3~cm) placed just upstream.
The beam rate is about 1~k particles/s and the trigger rate is about 300~triggers/s.
The signals from two trigger PMTs and MPPC are recorded using the CAEN DT5730 digitizer.

\subsection{Light yield}
\begin{figure}[tb]
    \centering
    \includegraphics[width=10cm]{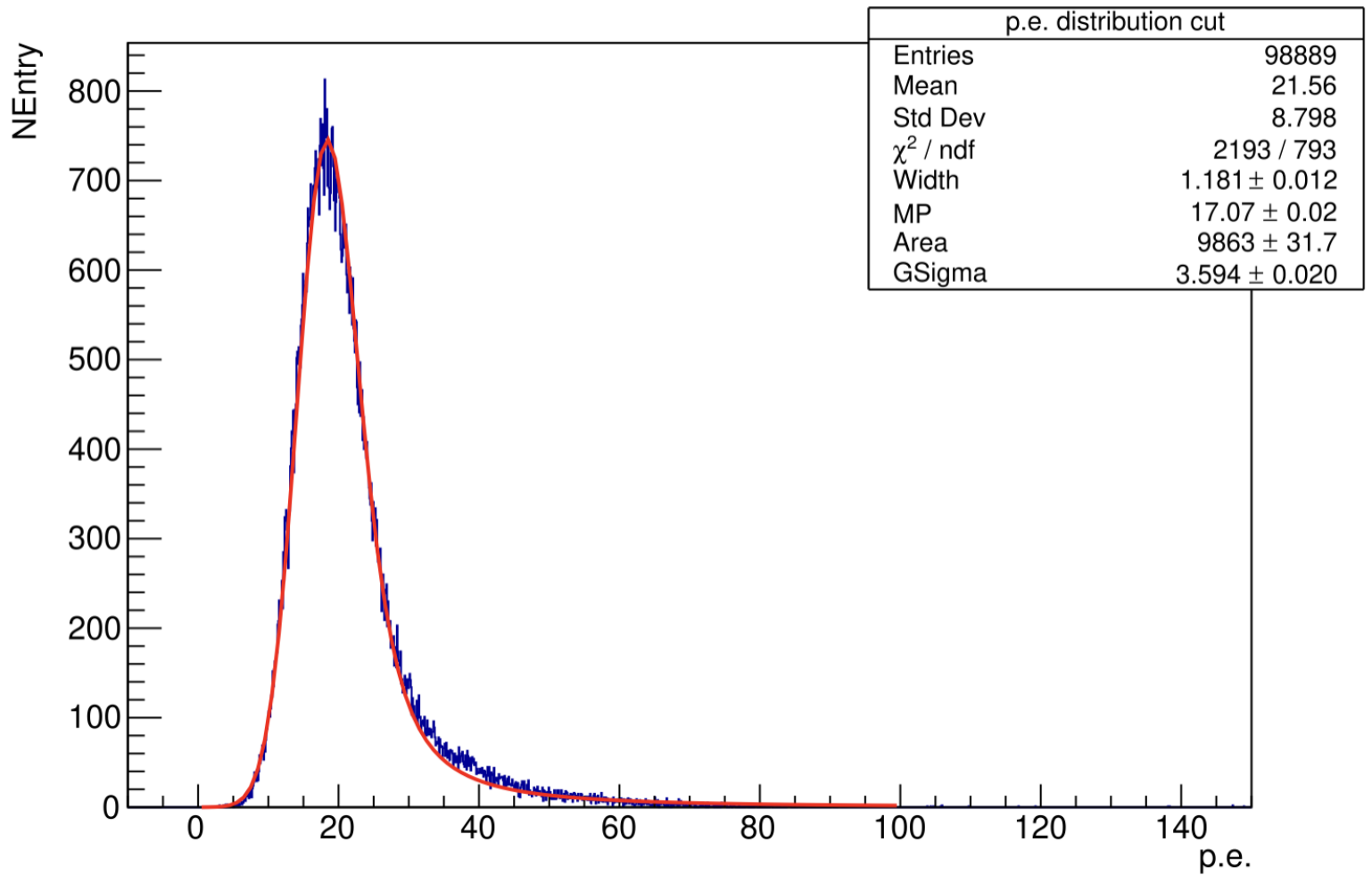}
    \caption{An example of the light yield distribution.}
    \label{fig:LightYieldDistribution}
\end{figure}

The light yield is measured with different combinations of the WLS fibers and plastic scintillators.
An example of the light yield distribution is shown in Fig.~\ref{fig:LightYieldDistribution}.
The number of events with higher light yield is larger than expected.
One possible reason is that the events that contain two electrons are included in our sample.
To reject such events, we require the signal of the trigger scintillator to be consistent with that of the single electron.
The distribution is fitted with the convolution function of a Landau distribution and a Gaussian function.

As shown in Fig.~\ref{fig:LightYieldDistribution}, the fitting uncertainties are 0.1\%.
The uncertainty from the rejection of double electron events is estimated from the difference between with and without the rejection to be 0.6\%.
The uncertainty of electron path length is estimated to be less than 0.1\%.
The variation of coupling between the fiber and scintillator is another source of uncertainty.
We measure Y-11 five times and estimate the uncertainty to be 1.0\%.
The effect from the distance between the scintillator and the MPPC is 0.1\%.
The temperature changes from 24.2$^\circ$C to 25.0$^\circ$C during the measurement, corresponding to $\pm0.4$\% change of the signal.
Adding all uncertainties in quadrature, the total uncertainty of light yield is estimated to be 1.2\%.

The measured light yields for each combination of scintillators and fibers,
defined as the most probable values of the Landau distribution,
are summarized in Table~\ref{tab:LightYield_Results1}.
\begin{table}[tb]
    \centering
    \begin{tabular}{cc|c}
        \hline
        Scintillator & Fiber & Light yield (p.e.) \\
        \hline \hline
        \multirow{4}{*}{EJ-200} & Y-11 & $17.12 \pm 0.21$ \\
               & YS-2 & $14.15 \pm 0.17$ \\
               & YS-4 & $12.13 \pm 0.15$ \\
               & YS-6 & $10.56 \pm 0.13$ \\
        \hline
        \multirow{4}{*}{EJ-204} & Y-11 & $17.84 \pm 0.21$ \\
               & YS-2 & $17.19 \pm 0.21$ \\
               & YS-4 & $15.71 \pm 0.19$ \\
               & YS-6 & $13.89 \pm 0.17$ \\
        \hline
    \end{tabular}
    \caption{Measured light yield (most probable value) for each combination of scintillator and WLS fiber.}
    \label{tab:LightYield_Results1}
\end{table}
With both scintillators, the Y-11 has the largest light yield.
Using EJ-204, the difference between the Y-11 and the YS series becomes smaller.
One of the reasons could be the matching of wavelengths between the emission from a scintillator and absorption by a fiber.

\subsection{Time resolution with plastic scintillators}

\begin{figure}[tb]
    \centering
    \subfloat[Y-11]{
        \includegraphics[width = 6.5cm]{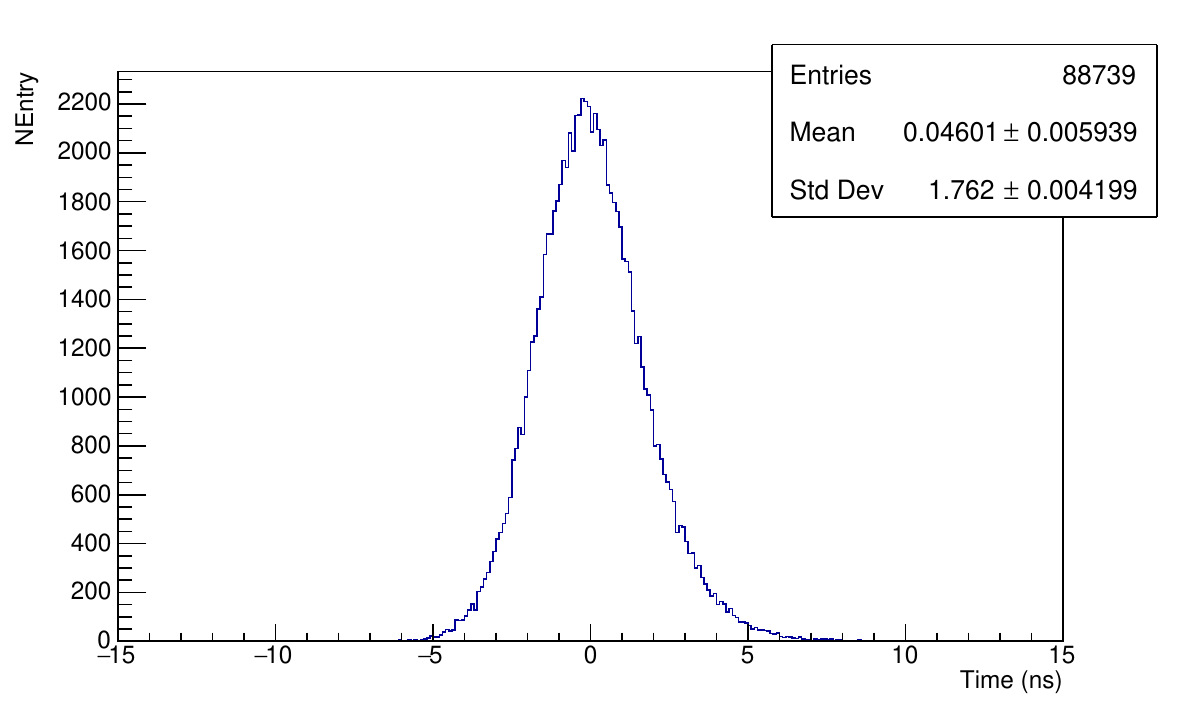}
        \label{fig:TimeResolutionResults_Y11}
    }
    \quad
    \subfloat[YS-2]{
        \includegraphics[width = 6.5cm]{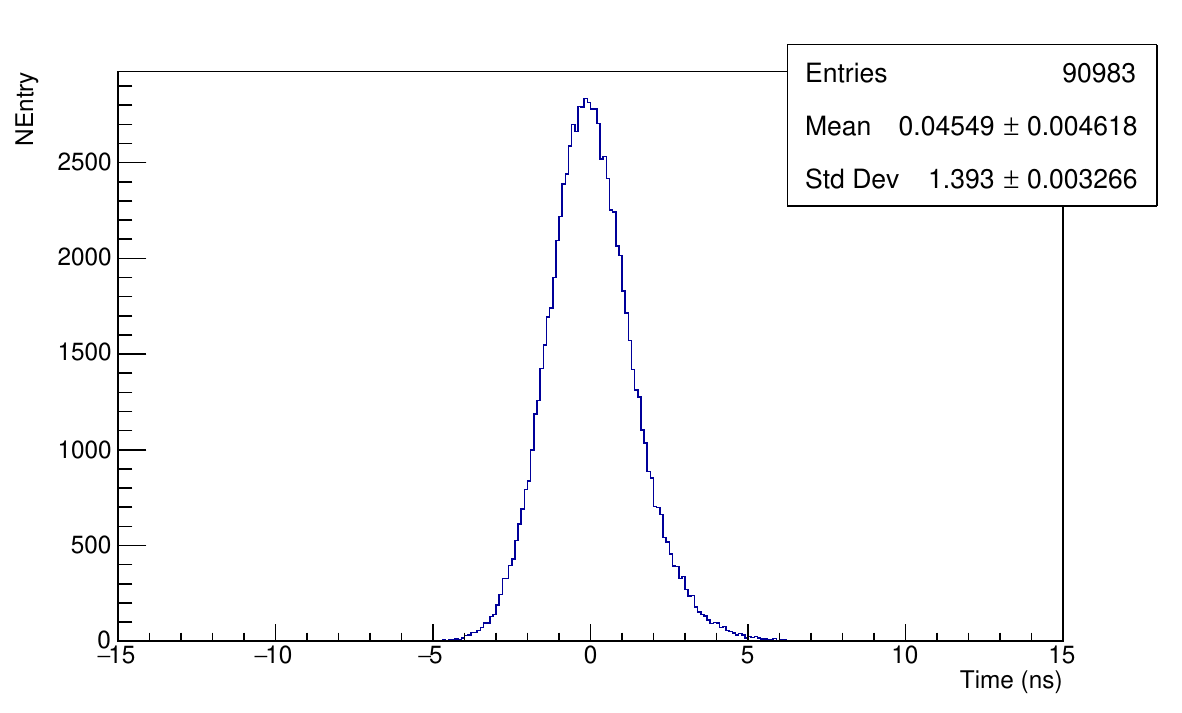}
        \label{fig:TimeResolutionResults_YS2}
    }
    \\
    \subfloat[YS-4]{
        \includegraphics[width = 6.5cm]{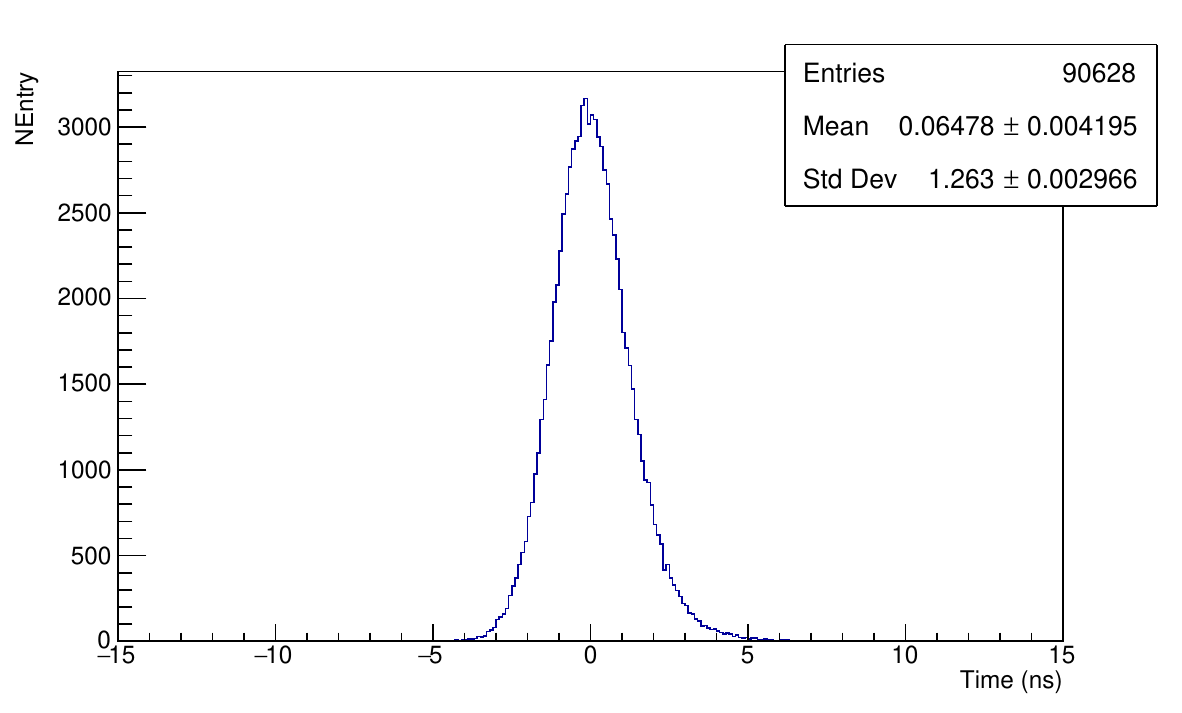}
        \label{fig:TimeResolutionResults_YS4}
    }
    \quad
    \subfloat[YS-6]{
        \includegraphics[width = 6.5cm]{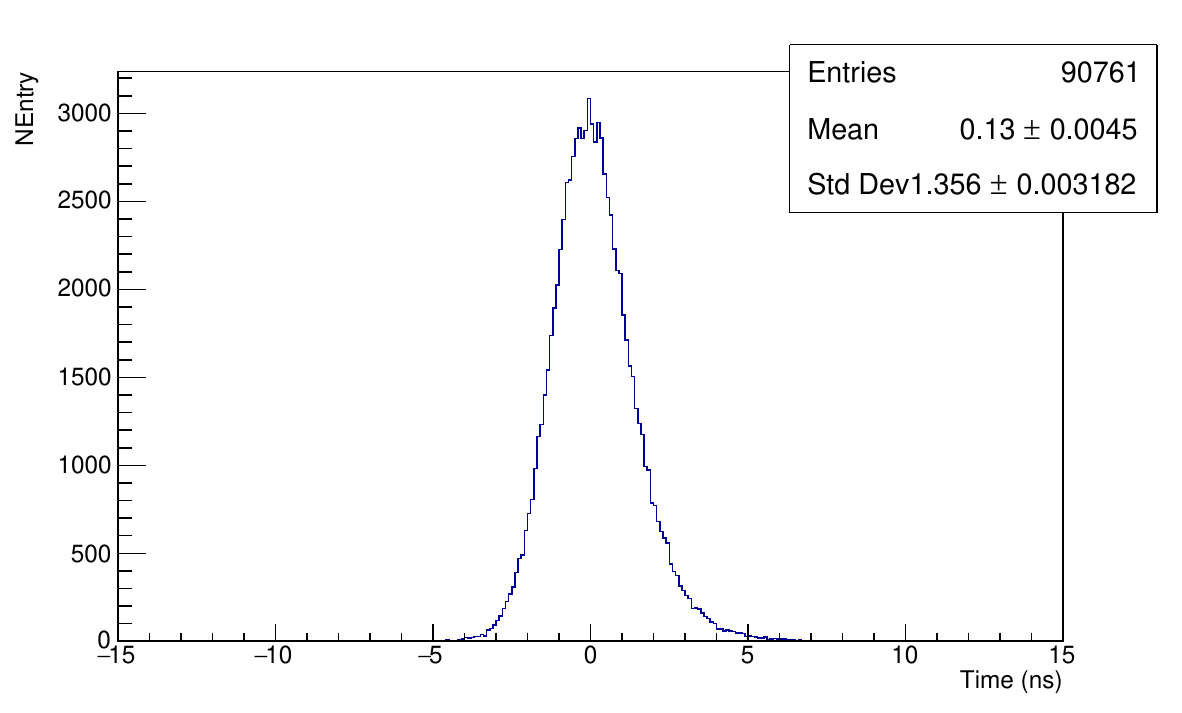}
        \label{fig:TimeResolutionResults_YS6}
    }
    \caption{Timing distributions for EJ-200.}
    \label{fig:TimeResolutionResults}
\end{figure}

We measure the time resolution of each combination of a scintillator and a fiber using the same data set as the light yield measurement.
The trigger time of an event is determined with the average of the signals of two PMTs attached to the trigger scintillator.
A template of the PMT signal is made from the waveforms of the same PMT recorded with a 2~GS/s oscilloscope.
We fit the digitizer waveform (500 MS/s) with the template to determine the timing of the PMT signal.
The standard deviation of the time difference between the two PMT signals is 0.25~ns.
Assuming that the time resolutions of two trigger PMTs are identical, the time resolution of the reference time is 0.13~ns.
The timing of the MPPC signal is defined by the same method as the measurement with the laser.
The time-walk effect is corrected based on the timing dependence on the observed light yield. 
The timing distributions for the four types of fibers with the EJ-200 scintillator are shown in Fig.~\ref{fig:TimeResolutionResults}.

\begin{table}[tb]
    \centering
    \begin{tabular}{cc|c}
        \hline
        Scintillator & Fiber & Timing resolution (ns) \\
        \hline \hline
        \multirow{4}{*}{EJ-200} & Y-11 & $1.77 \pm 0.04$ \\
               & YS-2 & $1.39 \pm 0.04$ \\
               & YS-4 & $1.26 \pm 0.04$ \\
               & YS-6 & $1.36 \pm 0.04$ \\
        \hline
        \multirow{4}{*}{EJ-204} & Y-11 & $1.63 \pm 0.04$ \\
               & YS-2 & $1.24 \pm 0.04$ \\
               & YS-4 & $1.07 \pm 0.04$ \\
               & YS-6 & $1.12 \pm 0.04$ \\
        \hline
    \end{tabular}
    \caption{Measured time resolution (standard deviation) of each combination.}
    \label{tab:TimeResolution_Results1}
\end{table}

The measured time resolutions, defined as the standard deviation of each distribution, are summarized in Table~\ref{tab:TimeResolution_Results1}.
We consider four sources of systematic uncertainties.
To check the reproducibility of the measurement, we measure the time resolution five times using Y-11 fibers.
Taking half of the maximum difference, 0.035~ns is assigned as the systematic uncertainty.
The uncertainty of time-walk correction is estimated by taking half of the difference between before and after the correction and found to be at most 0.012~ns.
The uncertainty of the MPPC timing determination is estimated by changing the number of points for the fitting and is estimated to be less than 0.01~ns.
The resolution of the trigger timing is 0.13~ns.
Taking half of the difference between with and without subtracting the contribution of trigger timing resolution, the uncertainty of the reference time resolution is evaluated to be less than 0.004~ns.
The dominant uncertainty is the reproducibility of the measurement.
Adding all the uncertainties in quadrature, the total uncertainty of the time resolution is estimated to be 0.04~ns.

Even though the light yield is smaller, the time resolution of the YS series is still better than that of the Y-11.
Among the YS series, YS-4 has the best time resolution although the difference is not significant compared to the measurement uncertainties.
For the YS series, the decay time of the scintillator is not negligible.
The decay times of the EJ-200 and EJ-204 (Table~\ref{tab:Scintillators}) are larger than the decay time of the YS-6.
Given this fact, these results are not inconsistent with the results of the decay time measurement using a laser.

\section{Conclusions}
The time resolution of the YS series is confirmed to be better than that of Y-11, both with laser light and in combination with plastic scintillators.
The attenuation length of the YS series is measured to be comparable to that of Y-11.
When combined with the plastic scintillators EJ-200 and EJ-204, the YS series have better time resolution than Y-11, with light yields of 60--100\% of Y-11.

\section*{Acknowledgments}

We thank Kuraray for providing samples of WLS fibers.
We acknowledge the support by KEK ITDC.
The measurement at the KEK PF-AR Test Beam Line was performed under the user program (2023ARTBL003).
This work was supported by JSPS KAKENHI Grant Number JP20H00149.
S.~K.\ is supported by the WINGS-QSTEP fellowship program of The University of Tokyo.

\bibliographystyle{ptephy}
\bibliography{bibliography}

\vspace{0.2cm}
\noindent

\let\doi\relax

\end{document}